# Mimicking synaptic plasticity with wedged Pt/Co/Pt spin-orbit torque device

Shiwei Chen[1†], Rahul Mishra[2,3†], Huanjian Chen[1], Hyunsoo Yang[2], Xuepeng Qiu[1*]

*[1]Shanghai Key Laboratory of Special Artificial Microstructure Materials and Technology and School of Physics Science and Engineering, Tongji University, Shanghai 200092, China*

*[2]Department of Electrical and Computer Engineering, National University of Singapore, 117576, Singapore*

*[3]Centre for Applied Research in Electronics, Indian Institute of Technology Delhi, New Delhi 110016, India*

[†] S. Chen and R. Mishra contributed equally to this work.

[*] Authors to whom correspondence should be addressed: xpqiu@tongji.edu.cn

**Abstract**

We fabricated a wedge shape of Pt/Co/Pt device with perpendicular magnetic anisotropy and manifested that the Co magnetization can be solely switched by spin-orbit torque without any magnetic field. Similar to the synaptic weight, we observed that the state of Co magnetization (presented by the anomalous Hall resistance $R_H$) of wedged Pt/Co/Pt device can be tuned continuously with a large number of nonvolatile levels by applied pulse currents. Furthermore, we have studied the synaptic plasticity of wedged Pt/Co/Pt device, including the excitatory postsynaptic potentials (EPSP) or inhibitory postsynaptic potentials (IPSP) and spiking-time-dependent plasticity (STDP). The work elucidates the promise of wedged Pt/Co/Pt device as a candidate for new type of artificial synaptic devices induced by spin current and paving a substantial pathway towards the combination of spintronics and synaptic devices.

Learning and memory in the human brain occur through dynamic changes in the connection or synaptic strength between interconnected networks of biological neurons. This biological network of neuron and synapses i.e. biological neural network is much more efficient than conventional computational systems in learning and performing pattern recognition tasks.[1-2] Inspired by the biological neural network, the field of neuromorphic computing, which has one of the aims to emulate synaptic functionalities by electronic device, has attracted significant interest.[3-5] However, traditional silicon-based complementary metal-oxide-semiconductor (CMOS) circuits use tens of transistors to mimic synaptic activity,[6] which requires a very large area and consumes a significant amount of power, hindering efficient scaling.[7] Therefore, designing artificial synapses with a single solid-state device is essential for effective hardware implementation of neural networks for brain-inspired computing.[8]

The demonstration of synapse-like electronic devices are mainly focused on nonvolatile charge memory technologies, including the phase change memory,[9-11] resistive switching memory,[12-13] ferroelectric memory,[14-15] magnetoelectric coupled memtranstors,[16] and electrolyte-gated three terminal transistors.[17-21] These devices mostly involve manipulating ions and/or electrons/holes and modulating the resulting multistate conductivity which represents the synaptic weight. In contrast with these non-volatile charge memory technologies, spintronic devices rely on the spin degree of freedom of electrons.[22-24] Spin devices such as spin-transfer torque (STT) device[25] and spin-orbit torque (SOT) device[26-27] are considered as a viable replacement of charge-based computing and memory elements due to their relatively low power, ultrafast operation and larger endurance. Due to the above reasons, spin devices also offer an ideal platform for alternate computing methodologies, as also demonstrated by the recent experimental works.[28-32] Among the spintronic device candidates, the SOT devices offer a simpler design compared to the STT ones. Using the SOT devices, although there have been stride towards implementing biological synapsed and neural functionalities,[33-35] the synaptic implementation is mostly limited to the demonstration of multilevel programming capability.[36-38] Besides, a small symmetry-breaking bias field, provided by either an external magnetic field or a unidirectional in-plane effective field via interlayer exchange coupling or exchange bias,[39-40] is needed for SOT device switching,[27] which is impractical in scaled designs.

In this paper, we report magnetic field-free SOT-induced nonvolatile multi-states behaviors in a wedged shape Pt/Co/Pt heterojunction. The anomalous Hall resistance ($R_H$) which represents the synaptic weight of the device can be tuned in an analog and nonvolatile fashion. Using Kerr microscope imaging, we show that a slowly propagating domain wall yields a broad range of intermediate resistive or magnetic states. The $R_H$ can be controlled by the pulsed current characteristics which include its amplitude, duration, and repetition number. This is similar to the stimulus characteristic dependent modulation of the biological synapse. The wedge-based spin device demonstrates many functionalities of biological synapse which include excitatory postsynaptic potential (EPSP), inhibitory postsynaptic potential (IPSP) and spike timing-dependent plasticity (STDP), important for developing neuromorphic computing.

The uniform sample substrate/Pt (1.5)/Co (0.6)/Pt (2.5)/SiO$_2$ (3) and the wedged sample substrate/Pt (1.5)/Co (0.6)/Pt (wedge)/SiO$_2$ (3) (nominal thicknesses in nm) were grown on thermally oxidized Si substrates by ultrahigh vacuum magnetron sputtering at room temperature. Wedge deposition was done by turning off the sample holder rotation during the sputtering of the top Pt layer. The nominal thickness of top Pt layer at the center is ~ 1.55 nm for the wedged sample. Note that the bottom Pt and Co layers are flat with respect to the substrate and only the top Pt layer is wedged. SiO$_2$ was deposited on top of the multilayers to prevent oxidation of the samples. Afterwards, the film was patterned into 8 μm wide Hall bars using lithography and Ar-ion milling techniques.

The layer structure and thickness gradient of the wedge sample and its relative orientation to the applied current during measurement are shown in Figure 1(a). The anomalous Hall resistance ($R_H$) as a function of external magnetic field ($H_z$) applied along the z axis is shown in Fig. 1(b). The anomalous Hall loops have a square shape demonstrating good perpendicular magnetic anisotropy (PMA) of the Co layer in both samples. Then, we measure the current-induced magnetization switching in the absence of any external magnetic field for the uniform and the wedged devices. The results are shown in Fig. 1(c). It should be noted that after the application of each current pulse a low amplitude $I$ (1 mA) was applied to evaluate the device state via $R_H$. As shown in Fig. 1(c), we observe a reversible field-free magnetization switching in the Pt/Co/Pt heterojunction prepared by wedged deposition, but no current-induced magnetization switching (subtracting the value of $R_H$ of uniform Pt/Co/Pt device) happens in the uniform Pt/Co/Pt heterojunction. Furthermore, we measured the current induced SOT switching with a fixed -50 Oe assist magnetic field ($H_x$). Deterministic magnetization switching can be achieved by currents for both samples as shown in Fig. 1(d). Similar to the previous results,[41-42] the anomalous Hall resistance of Pt/Co/Pt(uniform) device rapidly changes when the pulse current is larger than the critical switching current. However, for the wedged sample the anomalous Hall resistance increases gradually with the increasing pulse current before it reaching saturation state. It should be noted that while a $R_H$ vs. $I$ curve for a conventional SOT device has two stable binary states and a sharp switching trajectory, the wedged SOT device exhibits a gradual switching behavior characterized by multiple nonvolatile magnetic states.

The deterministic SOT switching of perpendicular magnetization without external magnetic field requires the symmetry breaking between the $\pm z$ magnetic states. Here, the wedged thickness of Pt layer breaks the mirror symmetry of the $xz$-plane, which allows for the emergence of an additional spin torque whose direction depends on the current polarity.[43-45] The symmetry breaking that enabling the field-free switching is not only restricted to the heavy metal layer but also has been demonstrated with the ferromagnet layer.[46] Either an spatially varying magnetization or anisotropy of ferromagnet is also capable to produce the field-free switching behaviours which further supports the mirror symmetry braking scenartio of our results.

To quantitatively measure the effective magnetic field ($H_{eff}$) in the wedged Pt/Co/Pt heterojunction, we investigate the current induced switching of the magnetization under different in-plane assisting magnetic fields. As shown in Fig. 1(e), with sweeping $H_x$ from -200 to -25 Oe applied in the current direction, a clockwise current-induced magnetization switching was observed. With $H_x$ = -18 Oe applied, there is no switching. With further increasing the external magnetic field, the deterministic current-induced magnetization switching was observed again. However, an opposite sign (anticlockwise) of current-induced magnetization switching was observed for $H_x$ varying from -12 to 200 Oe. This result demonstrates that the effective in-plane bias field is about 18 Oe. Based on the previous investigation, this effective in-plane bias field is a possible origin of the multiple nonvolatile magnetic state characteristics. The presented numerical simulations and experimental results reveal that the transition from binary magnetic states to multiple magnetic states derives from the tuning the efficiency of the spin–orbit torque to drive domain wall motion in a pinning potential by the in-plane field component.[34]

Next, we investigate the change in $R_H$ induced by pulse currents with various sweeping ranges. Magnetization switching was performed with in-plane pulsed current injection as schematically shown in Figure 2(a). Figure 2(b) shows a plot of the $R_H$ measured with the pulse current scanning from −30 mA to several different $I_{max}$ and then back to −30 mA, where the $I_{max}$ varies from 18 to +30 mA. For different values of $I_{max}$, the device attains various intermediate magnetization states. These intermediate magnetization states were found to be non-volatile. To gain an insight into the microscopic mechanisms responsible for this multi-state effect, we investigate the correlation between the Hall resistance value and the magnetic domain configuration which is imaged by a polar Kerr microscope with an LED light source (TuoTuo

Technology). We have collected magnetic domain patterns after setting the device into different magnetic states. The corresponding Kerr microscope images of the Hall bar during its down-to-up magnetic switching processes are shown in Fig. 2(c). At the beginning of the down-to-up switching process with the threshold pulse magnitude $I$ = 18 mA, nucleation of a small domain is found to appear at the edge of the Hall bar, as shown by the light area in Fig. 2c (1). The domain wall propagates across the Hall bar with the increasing pulse current (Figures 2c (1) - 2c (5)), and finally fills up the Hall bar area at a finishing switching current $I$ = 30 mA (Figure 2c (6)). Thus, in wedged Pt/Co/Pt devices, a nonvolatile multi-state behavior can be devised by controlling the nucleation and growth of magnetic domain. Next, we discuss the correlation between the functionality of the wedged device and a biological synapse.

A biological synapse is a conjunction of two neuron cells, named pre-synaptic neuron and post- synaptic neuron, as shown in Fig. 3(a). When a positive electrical signal arrives in a pre-synaptic neuron, it opens the voltage-gated calcium channels and generates a $Ca^{2+}$ flow into the neuron. A rapidly increasing concentration of $Ca^{2+}$ triggers the release of neurotransmitters, which dock with receptors onto the postsynaptic neuron. If enough neurotransmitters dock with receptors, an electric stimulus responds in postsynaptic neuron and generates excitatory postsynaptic potentials (EPSP) or inhibitory postsynaptic potentials (IPSP). While an EPSP increases the membrane potential of a neuron, an IPSP decreases it. In this way, information from the pre-synaptic neuron is transmitted through the synapse to the post-neuron.[47] The strength of the connection between neurons is defined as synaptic weight. The property of synapse to modulate its weight according to the input stimulus is known as "synaptic plasticity" and represents the foundation of learning and memory in neural systems.

As shown in Fig. 3(b), the spin current/spin accumulation generated by spin-Hall/Rashba effect in Pt layers diffuses into the Co layer while the in-plane current is applied on the wedged Pt/Co/Pt heterojunction. This spin current exerts spin-orbit torques (SOTs) on the Co magnetization and modulate Co magnetization orientation (i.e. increasing/decreasing the value of $R_H$). Therefore, the applied pulse current plays the roles of the pre- and post-spikes that act on our artificial synapse (wedged Pt/Co/Pt device) to modulate $R_H$ (i.e., synaptic weight), which is assessed by the detection of $R_H$ (i.e., EPSP/IPSP).

Next, we show basic synapse characteristic, such as potentiation, depression, spike-width dependent plasticity and timing-dependent plasticity, using wedged Pt/Co/Pt artificial synapse. Fig. 3(c) shows the increasing and decreasing of $R_H$ by applying sequences of 30 pulses with a constant magnitude (lower panel) and duration of 0.1 ms. This indicates the potentiation and depression of the synaptic weight. Subsequently, with increasing the pulse number, $R_H$ increases gradually before saturating. Reversing the direction of the pulse current, $R_H$ begins to decreases and saturates to a low $R_H$ state. An important characteristic of the wedged device is the repeatability of the achieved state as shown during several cycles of potentiation and depression in Fig. 3(d). By both amplitude and polarities of current pulses, multi-magnetic-states with different $R_H$ values can be obtained. In biological systems, the synaptic plasticity can be divided into short-term plasticity and long-term plasticity according to the retention time of the synaptic weight, corresponding to the short-term memory and long-term memory of brain, respectively.[1, 48-49] Long-term potentiation is widely considered as the mechanism that underlies learning and memory in biological systems.[50] As shown in Fig. 3(d), we can observe that $R_H$ value is stable at a high-resistance state under pulse current of 0 mA after setting a potentiation state with positive pulse currents in the wedged Pt/Co/Pt artificial synapse, and vice versa. Therefore, the EPSP and IPSP of wedged Pt/Co/Pt artificial synapse can be classed as long-term plasticity, corresponding to the long-term potentiation (LTP) and long-term depression (LTD), respectively. Fig. 3(e) shows a typical potentiation response of this artificial synapse triggered by pulse currents of a fixed magnitude (23 mA) but variable pulses width. Before the application of sequence, an initialization pulse with a magnitude of -30 mA was applied to set a negative maximum $R_H$ state. For all the measurements with different pulses widths, $R_H$ increases with increasing the pulse number, which corresponds to an increase of the synaptic weight. For a larger width of pulses, the synaptic weight more rapidly changes and increases by a larger amount compared to the case for which the pulses width is shorter. This pulse-width dependent plasticity is comparable to the spike-width dependent plasticity of the biological synapse which shows a larger weight modulation when it receives a stronger stimulus.

A well-known mechanism for synaptic plasticity is Hebbian learning, which suggests that the synaptic weight is modulated in accordance to neural activities in pre-neuron and post-neuron cells.[51] In this form, one of the most important learning rules is spiking-time-dependent plasticity (STDP).[52] STDP establishes the synaptic weight adjustment according to the timing of the fired spikes by connected neurons. According to this rule, the synapse potentiates (increase of $R_H$) if pre-synaptic spikes precede postsynaptic spikes, and the synapse depresses (decrease of $R_H$) if postsynaptic spike precedes pre-synaptic spike. More specifically, the precise pre- and post-spike timing window controls the sign and magnitude of synaptic weight modulation. A biological synapse realizes such plastic functions by regulating the ion concentrations inside it. For the case of wedged Pt/Co/Pt device, STDP is emulated by engineering the pre- and post-spike superimpositions. Before the application of each single pulse, the $R_H$ was reset to zero. When the amplitude of pulse current is larger than the absolute value of the current threshold ($I_{th}$), an increase or a decrease of $R_H$ can be induced by the single pulse and higher amplitude pulse brings a greater change in $R_H$ until the current approaches saturation, as shown in Fig. 4(a).

Based on this $I$ dependence of $R_H$ under a single current pulse, the spike is shaped as a biological spike and designed for the demonstration of STDP in our device, as shown in the Fig. 4(b). It should be noted that the pre-spike and post-spike share the same waveform but are opposite in polarity. In this way, although the current density of a single pre- or post-spike never exceeds $I_{th}$, the overlapped waveform will be above the $I_{th}$, whose sign and magnitude depend on the relative arrival of pre and postsynaptic spikes ($\Delta t = t_{pre} - t_{post}$), as shown in the bottom panel of Fig. 4(b). The superimposition between the pre-spike and the post-spike ($I_{pre} - I_{post}$) defines the net programming current applied on the synapse. By repeating the pulse scheme with different $\Delta t$ in the range from −20 to 20 ms, the STDP curve is obtained, as shown in the Fig. 4(c). While the STDP curve is not an ideal exponential, the synaptic weight modulation is inversely proportional to $\Delta t$, which is sufficient to enable Hebbian learning.

In summary, we have demonstrated a spin based multistate artificial synapse with wedged Pt/Co/Pt spin-orbit torque devices. The multi magnetic state feature is confirmed to be dependent on the wedged Pt layer. The ability to set multiple magnetic states, depending on the pulse number and magnitude, enables spintronic implementations of synapse functionalities including EPSP, IPSP and STDP. Our results suggest a practical way toward spintronic synaptic emulation using current controlled multistates in perpendicular magnetic materials with built-in equivalent in-plane magnetic fields, as artificial synapses for neuromorphic computing.

This work was supported by the National Key R & D Program of China Grand No. 2017YFA0303202 and 2017YFA0305300, the National Science Foundation of China Grant Nos. 52022069, 11974260, 11674246, 51501131, 51671147, 11874283, 51801152, and 11774064, Natural Science Foundation of Shanghai Grant No. 19ZR1478700, the Fundamental Research Funds for the Central Universities, and SpOT-LITE program (A*STAR grant, A18A6b0057) through RIE2020 funds.

The data that support the findings of this study are available from the corresponding author upon reasonable request.

# References:

1. L. F. Abbott and S. B. Nelson, Nat. Neurosci. **3**, 1178 (2000).
2. W. Gerstner, H. Sprekeler, and G. Deco, Science **338**, 60 (2012).
3. D. Kuzum, S. Yu, and H. S. Wong, Nanotechnology **24**, 382001 (2013).
4. M. Prezioso, F. Merrikh-Bayat, B. D. Hoskins, G. C. Adam, K. K. Likharev, and D. B. Strukov, Nature **521**, 61 (2015).
5. M. A. Zidan, J. P. Strachan, and W. D. Lu, Nat. Electron. **1**, 22 (2018).
6. E. Chicca, D. Badoni, V. Dante, M. D'Andreagiovanni, G. Salina, L. Carota, S. Fusi, and P. Del Giudice, IEEE Trans. Neural. Netw. **14**, 1297 (2003).
7. G. Indiveri and S. C. Liu, Proc. IEEE **103**, 1379 (2015).
8. D. S. Jeong and C. S. Hwang, Adv. Mater. **30**, 1704729 (2018).
9. D. Kuzum, R. G. Jeyasingh, B. Lee, and H. S. Wong, Nano Lett. **12**, 2179 (2012).
10. J. J. Yang, M. D. Pickett, X. Li, D. A. Ohlberg, D. R. Stewart, and R. S. Williams, Nat. Nanotechnol. **3**, 429 (2008).
11. A. Athmanathan, M. Stanisavljevic, N. Papandreou, H. Pozidis, and E. Eleftheriou, IEEE J. Emerg. Sel. Top. Circuits Syst. **6**, 87 (2016).
12. Z. R. Wang, S. Joshi, S. E. Savelev, H. Jiang, R. Midya, P. Lin, M. Hu, N. Ge, J. P. Strachan, Z. Li, Q. Wu, M. Barnell, G. L. Li, H. L. Xin, R. S. Williams, Q. F. Xia, and J. J. Yang, Nat. Mater. **16**, 101 (2017).
13. S. H. Jo, T. Chang, I. Ebong, B. B. Bhadviya, P. Mazumder, and W. Lu, Nano Lett. **10**, 1297 (2010).
14. N. Setter, D. Damjanovic, L. Eng, G. Fox, S. Gevorgian, S. Hong, A. Kingon, H. Kohlstedt, N. Y. Park, G. B. Stephenson, I. Stolitchnov, A. K. Tagantsev, D. V. Taylor, T. Yamada, and S. Streiffer, J. Appl. Phys. **100**, 051606 (2006).
15. A. Chanthbouala, V. Garcia, R. O. Cherifi, K. Bouzehouane, S. Fusil, X. Moya, S. Xavier, H. Yamada, C. Deranlot, N. D. Mathur, M. Bibes, A. Barthélémy, and J. Grollier, Nat. Mater. **11**, 860 (2012).
16. J. X. Shen, D. S. Shang, Y. S. Chai, S. G. Wang, B. G. Shen, and Y. Sun, Adv. Mater. **30**, 1706717 (2018).
17. J. Shi, S. D. Ha, Y. Zhou, F. Schoofs, and S. Ramanathan, **4**, 2676 (2013).
18. Y. Ren, J. Q. Yang, L. Zhou, J. Y. Mao, S. R. Zhang, Y. Zhou, and S. T. Han, Adv. Funct. Mater. **28**, 1805599 (2018).
19. Y. H. Liu, L. Q. Zhu, P. Feng, Y. Shi, and Q. Wan, Adv. Mater. **27**, 5599 (2015).
20. J. T. Yang, C. Ge, J. Y. Du, H. Y. Huang, M. He, C. Wang, H. B. Lu, G. Z. Yang, and K. J. Jin, Adv. Mater. **30**, 1801548 (2018).
21. R. Mishra, D. Kumar, and H. Yang, Phys. Rev. Appl. **11**, 054065 (2019).
22. F. Pulizzi, Nat. Mater. **11**, 367 (2012).
23. S. Manipatruni, D. E. Nikonov, and I. A. Young, Nat. Phys. **14**, 338 (2018).
24. I. Zutic, J. Fabian, and S. D. Sarma, Rev. Mod. Phys. **76**, 323 (2004).
25. F. J. Albert, J. A. Katine, R. A. Buhrman, and D. C. Ralph, Appl. Phys. Lett. **77**, 3809 (2000).
26. I. M. Miron, K. Garello, G. Gaudin, P.-J. Zermatten, M. V. Costache, S. Auffret, S. Bandiera, B. Rodmacq, A. Schuhl, and P. Gambardella, Nature **476**, 189 (2011).
27. L. Liu, O. J. Lee, T. J. Gudmundsen, D. C. Ralph, and R. A. Buhrman, Phys. Rev. Lett. **109**, 096602 (2012).

28  P. Krzysteczko, J. Munchenberger, M. Schafers, G. Reiss, and A. Thomas, Adv. Mater. **24**, 762 (2012).

29  J. Torrejon, M. Riou, F. A. Araujo, S. Tsunegi, G. Khalsa, D. Querlioz, P. Bortolotti, V. Cros, K. Yakushiji, A. Fukushima, H. Kubota, S. Y. Uasa, M. D. Stiles, and J. Grollier, Nature **547**, 428 (2017).

30  S. Lequeux, J. Sampaio, V. Cros, K. Yakushiji, A. Fukushima, R. Matsumoto, H. Kubota, S. Yuasa, and J. Grollier, Sci. Rep. **6**, 31510 (2016).

31  N. Locatelli, V. Cros, and J. Grollier, Nat. Mater. **13**, 11 (2014).

32  J. Grollier, D. Querlioz, and M. D. Stiles, Proc. IEEE Inst. Electr. Electron. Eng. **104**, 2024 (2016).

33  W. A. Borders, H. Akima, S. Fukami, S. Moriya, S. Kurihara, Y. Horio, S. Sato, and H. Ohno, Appl. Phys. Express **10**, 013007 (2017).

34  Y. Cao, A. Rushforth, Y. Sheng, H. Zheng, and K. Wang, Adv. Funct. Mater. **29**, 1808104 (2019).

35  J. Yun, Q. Bai, Z. Yan, M. Chang, J. Mao, Y. Zuo, D. Yang, L. Xi, and D. Xue, Adv. Funct. Mater. **30**, 1909092 (2020).

36  A. Sengupta and K. Roy, Appl. Phys. Express **11**, 030101 (2018).

37  A. Sengupta and K. Roy, IEEE Trans. Circuits Syst. I, **63**, 2267 (2016).

38  A. Sengupta, Y. Shim, and K. Roy, IEEE Trans. Biomed. Circuits. Syst. **10**, 1152 (2016).

39  Y.-C. Lau, D. Betto, K. Rode, J. M. D. Coey, and P. Stamenov, Nat. Nanotechnol. **11**, 758 (2016).

40  S. Fukami, C. Zhang, S. DuttaGupta, A. Kurenkov, and H. Ohno, Nat. Mater. **15**, 535 (2016).

41  X. Zhao, Y. Liu, D. Zhu, M. Sall, X. Zhang, H. Ma, J. Langer, B. Ocker, S. Jaiswal, G. Jakob, M. Kläui, W. Zhao, and D. Ravelosona, Appl. Phys. Lett. **116**, 242401 (2020)

42  S. Zhang, Y. Su, X. Li, R. Li, W. Tian, J. Hong, and L. You, Appl. Phys. Lett. **114**, 042401 (2019)

43  G. Yu, P. Upadhyaya, Y. Fan, J. G. Alzate, W. Jiang, K. L. Wong, S. Takei, S. A. Bender, L. Te Chang, Y. Jiang, M. Lang, J. Tang, Y. Wang, Y. Tserkovnyak, P. K. Amiri, and K. L. Wang, Nat. Nanotechnol. **9**, 548 (2014).

44  T. Chen, H. Chan, W. Liao, and C. Pai, Phys. Rev. Appl. **10**, 044038 (2018)

45  R. Chen, Q. Cui, L. Liao, Y. Zhu, R. Zhang, H. Bai, Y. Zhou, G. Xing, F. Pan, H. Yang, and C. Song, Nat. Commun. **12**, 3113 (2021).

46  H. Wu, J. Nance, S. A. Razavi, D. Lujan, B. Dai, Y. Liu, H. He, B. Cui, D. Wu, K. Wong, K. Sobotkiewich, X. Li, G. P. Carman, and K. L. Wang, Nano Lett. **21**, 515 (2021)

47  D. Purves, G. J. Augustine, D. Fitzpatrick, W. C. Hall, A.-S. LaMantia, J. O. McNamara, and S. M.Williams, *Neuroscience*, 3rd ed., Sinauer Associates, Sunderland, MA 2004.

48  T. Ohno, T. Hasegawa, T. Tsuruoka, K. Terabe, J. K. Gimzewski, and M. Aono, Nat. Mater. **10**, 591 (2011).

49  T. Chang, S. H. Jo, and W. Lu, Acs Nano **5**, 7669 (2011).

50  W. Xu, S. Y. Min, H. Hwang, and T. W. Lee, Sci. Adv. **2**, 1501326 (2016).

51  D. O. Hebb, Wiley Book Clin. Psychol. 62 (1949).

52  N. Caporale and Y. Dan, Annu. Rev. Neurosci. **31**, 25 (2008).

Figure 1

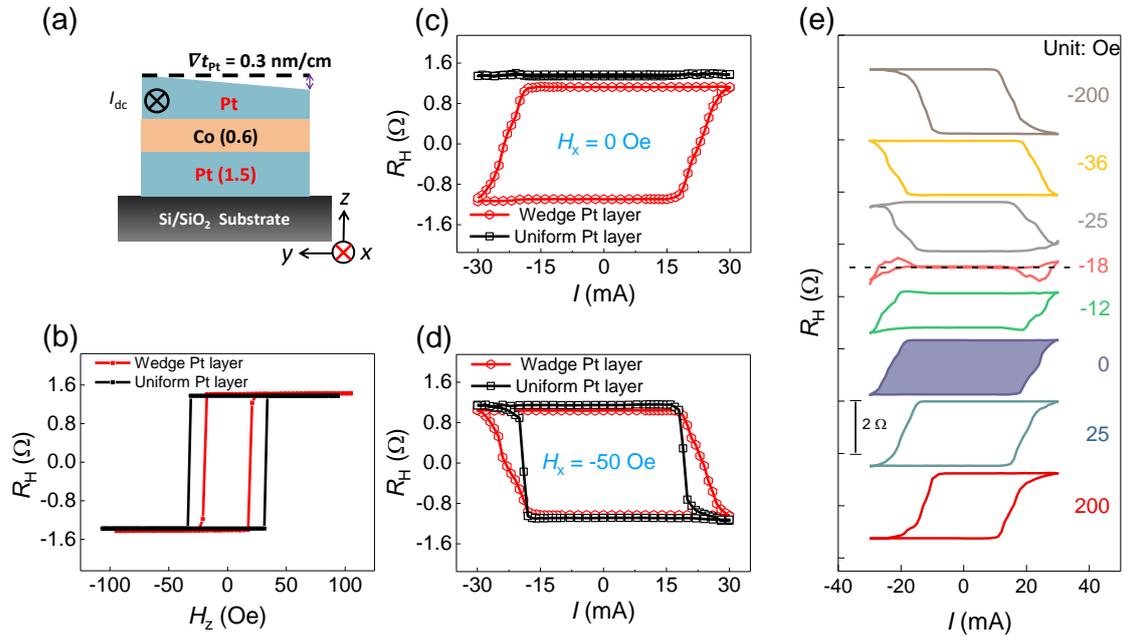

**FIG. 1.** (a) Layer structure of the wedge-deposited Pt sample (not to scale). (b) The anomalous Hall resistance ($R_H$) curves as functions of external field in the z direction for uniform Pt (black line) and wedge-deposited Pt (red line) Hall-bar samples. (c, d) Magnetization switching induced by the pulse current for uniform Pt (black line) and wedge-deposited Pt (red line) device (c) without external field and (d) with a fixed external magnetic field $H_x$ = -50 Oe. (e) The measured $R_H$ as a function of the pulse current at various applied fields for the wedge-deposited Pt/Co/Pt device.

Figure 2

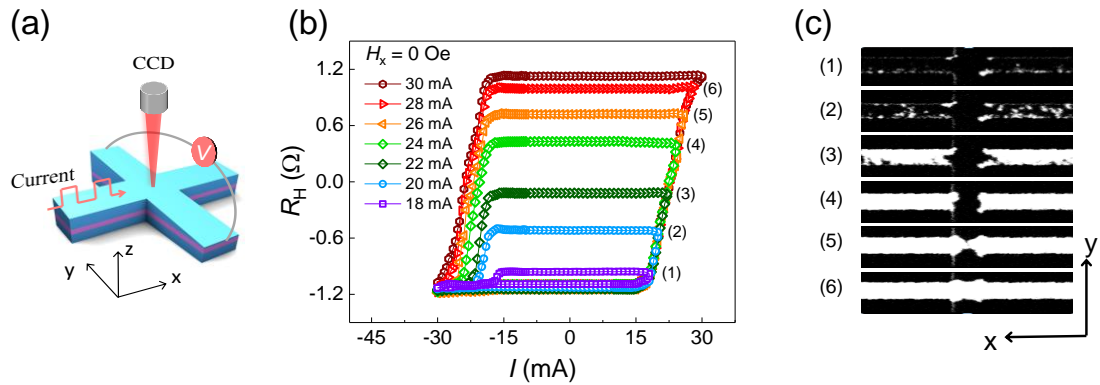

**FIG. 2.** (a) Schematic for measuring the current-induced magnetization switching, which was simultaneously recorded by Hall voltage measurements and Kerr microscopy imaging. (b) The measured $R_H$ as a function of pulse current with different maximum currents. A sequence of pulses with scanning magnitude from 30 mA to $-I_{MAX}$ and then back to 30 mA is applied. (c) Kerr microscope images during the down-to-up (black to white) switching process.

Figure 3

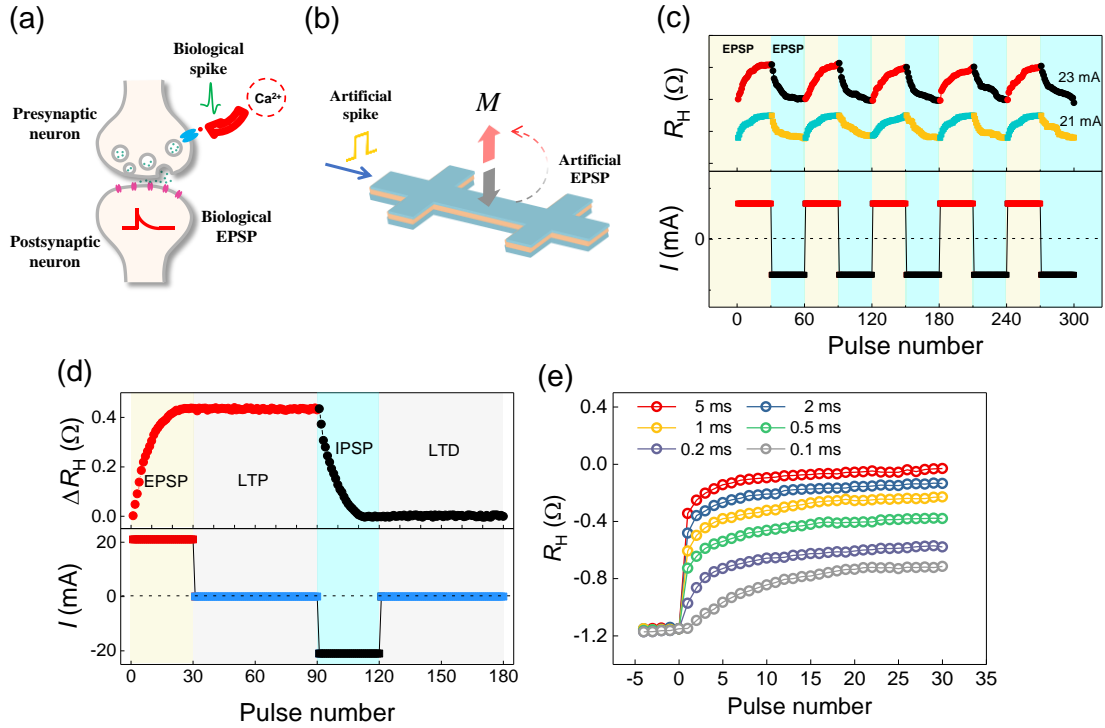

**FIG. 3.** (a) The information transmission between neurons occurs via propagation of action potentials through the axon and release of neurotransmitters. (b) The wedged Pt/Co/Pt artificial synapse transmits information by applying current pulses to change the Co magnetization to produce the anomalous Hall resistance $R_H$ variation. The $R_H$ plays a role of either EPSP (increase of $R_H$) or IPSP (decrease of $R_H$) in the postsynaptic axon. (c) The evolution of the EPSP/IPSP (i.e., $R_H$) by applying trains of pulse current (bottom part) with a constant amplitude. (d) Long-term potentiation and depression mimicked by a stimulus train with multiple pulses. (e) Evolution of the anomalous Hall resistance of the magnetic synapse under the application of current pulse of 23 mA magnitude. Various curves represent the device evolution for the different pulse widths.

Figure 4

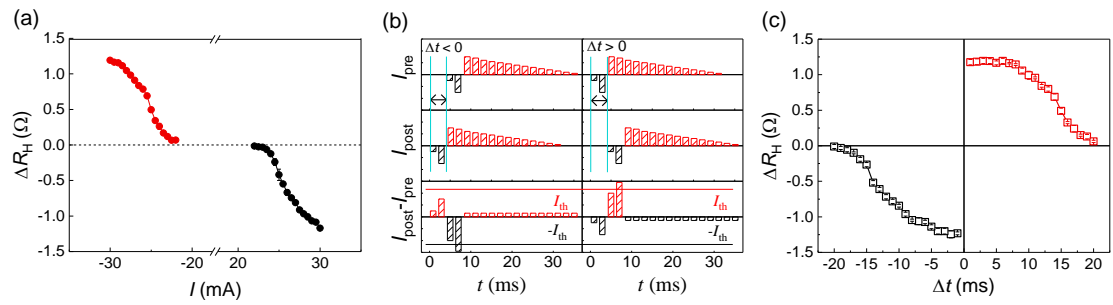

**FIG. 4.** (a) $\Delta R_H$ as a function of pulse current (initialized in the $R_H \approx 0$) when only one current pulse with a duration of 0.1 ms is applied. (b) The sequence of current pulse and (c) the corresponding experimental data showing the conventional STDP characteristic.

19